\documentclass{article}
\usepackage{spconf,amsmath,amssymb,url,bm}
\usepackage{tikz}
\usepackage{comment}
\usepackage{algorithmic}
\usepackage{algorithm}
\usepackage{booktabs}
\usepackage[
    backend=biber,
    style=ieee,
    maxbibnames=3,
    maxcitenames=3,
    doi=false,isbn=false,url=false,eprint=false
]{biblatex} 
\usepackage{multirow}
\usepackage[subrefformat=parens]{subcaption}

\usepackage{hyperref}
\hypersetup{
    colorlinks = false,
}

\DeclareSourcemap{
	\maps[datatype=bibtex, overwrite=true]{
		\map{
			\step[fieldsource=booktitle,
			match=\regexp{.*Interspeech.*},
			replace={Proc. INTERSPEECH}]
			\step[fieldsource=journal,
			match=\regexp{.*INTERSPEECH.*},
			replace={Proc. INTERSPEECH}]
			\step[fieldsource=booktitle,
			match=\regexp{.*ICASSP.*},
			replace={Proc. ICASSP}]
			\step[fieldsource=booktitle,
			match=\regexp{.*icassp_inpress.*},
			replace={ICASSP (in press)}]
			\step[fieldsource=booktitle,
			match=\regexp{.*International.*Conference.*on.*Acoustics,.*Speech.*and.*Signal.*Processing.*},
			replace={ICASSP}]
			\step[fieldsource=booktitle,
			match=\regexp{.*International.*Conference.*on.*Learning.*Representations.*},
			replace={ICLR}]
			\step[fieldsource=booktitle,
			match=\regexp{.*International.*Conference.*on.*Machine.*Learning.*},
			replace={Proc. ICML}]
			\step[fieldsource=booktitle,
			match=\regexp{.*Automatic.*Speech.*Recognition.*and.*Understanding.*},
			replace={Proc. ASRU}]
			\step[fieldsource=booktitle,
			match=\regexp{.*Spoken.*Language.*Technology.*},
			replace={Proc. SLT}]
			\step[fieldsource=booktitle,
			match=\regexp{.*Speech.*Synthesis.*Workshop.*},
			replace={Proc. SSW}]
			\step[fieldsource=booktitle,
			match=\regexp{.*workshop.*on.*speech.*synthesis.*},
			replace={Proc. SSW}]
			\step[fieldsource=booktitle,
			match=\regexp{.*Advances.*in.*neural.*information.*processing.*},
			replace={Proc. NIPS}]
			\step[fieldsource=booktitle,
			match=\regexp{.*Advances.*in.*Neural.*Information.*Processing.*},
			replace={Proc. NIPS}]
			\step[fieldsource=booktitle,
			match=\regexp{.*AAAI.*},
			replace={Proc. AAAI}]
			\step[fieldsource=booktitle,
			match=\regexp{.*Workshop.*on.* Applications.* of.* Signal.*Processing.*to.*Audio.*and.*Acoustics.*},
			replace={Proc. WASPAA}]
			\step[fieldsource=booktitle,
			match=\regexp{.*International.*Conference.*on.*Language.*Resources.*and.*Evaluation.*},
			replace={Proc. LREC}]
			\step[fieldsource=journal,
			match=\regexp{.*Spontaneous.*Speech.*Processing.*and.*Recognition},
			replace={Proc. SSPR}]
			\step[fieldsource=publisher,
			match=\regexp{.+},
			replace={{}}]
			\step[fieldsource=month,
			match=\regexp{.+},
			replace={{}}]
			\step[fieldsource=location,
			match=\regexp{.+},
			replace={{}}]
			\step[fieldsource=address,
			match=\regexp{.+},
			replace={{}}]
			\step[fieldsource=organization,
			match=\regexp{.+},
			replace={{}}]
			\step[fieldsource=doi,
			match=\regexp{.+},
			replace={{}}]
			\step[fieldsource=url,
			match=\regexp{.+},
			replace={{}}]
			\step[fieldsource=editor,
			match=\regexp{.+},
			replace={{}}]
		}
	}
} 

\title{Diversity-based core-set selection for text-to-speech\\with linguistic and acoustic features}

\name{Kentaro Seki, Shinnosuke Takamichi, Takaaki Saeki, and Hiroshi Saruwatari\thanks{This work is supported by JSPS KAKENHI 22H03639 and Moonshot R\&D Grant Number JPMJPS2011. We also appreciate Dong Yang of the University of Tokyo for his help.}}
\address{The University of Tokyo, Japan.}

\makeatletter
\newcommand{\figcaption}[1]{\def\@captype{figure}\caption{#1}}
\newcommand{\tblcaption}[1]{\def\@captype{table}\caption{#1}}
\makeatother

\begin{document}
\ninept
\maketitle
\setlength{\tabcolsep}{1mm} 
\setlength{\abovedisplayskip}{3pt} 
\setlength{\belowdisplayskip}{3pt} 


\begin{abstract} \vspace{-2mm}
This paper proposes a method for extracting a lightweight subset from a text-to-speech (TTS) corpus ensuring synthetic speech quality.
In recent years, methods have been proposed for constructing large-scale TTS corpora by collecting diverse data from massive sources such as audiobooks and YouTube.
Although these methods have gained significant attention for enhancing the expressive capabilities of TTS systems, they often prioritize collecting vast amounts of data without considering practical constraints like storage capacity and computation time in training, which limits the available data quantity. 
Consequently, the need arises to efficiently collect data within these volume constraints.
To address this, we propose a method for selecting the core subset~(known as \textit{core-set}) from a TTS corpus on the basis of a \textit{diversity metric}, which measures the degree to which a subset encompasses a wide range.
Experimental results demonstrate that our proposed method performs significantly better than the baseline phoneme-balanced data selection across language and corpus size.
\end{abstract} \vspace{-1.2mm}


\vspace{-2mm}
\section{Introduction}
\vspace{-2mm}
Although text-to-speech~(TTS) has achieved human-level naturalness in transforming text into speech waveform~\cite{Tacotron2}, its expressive capabilities do not yet match those of humans.
Previous studies have aimed to enhance TTS expressiveness by addressing aspects such as speaker identity control~\cite{arik2018neural}, emotional expression~\cite{li2021controllable}, and prosody representation~\cite{skerry2018towards}.
These studies predominantly adopt data-driven methods based on machine learning, and corpora with wider speaker or style variation are increasingly being anticipated.

To construct TTS corpora containing diverse data, previous studies have gathered data from vast sources such as audiobooks~\cite{LibriTTS} and YouTube~\cite{seki2023}.
These methodologies are predicated on the belief that collecting data from large-scale sources inherently results in a diverse dataset, and they utilize all available data. 
Consequently, recently released speech corpora~\cite{chen2021gigaspeech2, JTubeSpeech} comprise several thousand or even tens of thousands of hours of data.
In practice, however, limitations on storage capacity and the time required for learning impose constraints on the amount of available data, requiring datasets to be efficiently collected within these constraints~\cite{menon2020development,luo2021lightspeech}.
Given this context, TTS training in practical environments is expected to be achieved if data size can be reduced without compromising the quality of synthetic speech.

\begin{figure}[t]
  \centering
  \begin{minipage}{0.48\linewidth}
    \centering
    \includegraphics[width=1.\linewidth]{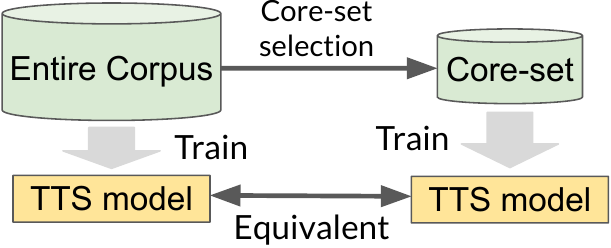}
    \vspace{-5mm}
    \subcaption{The core-set ideally has a learning effect equivalent to the entire dataset.}
    \label{fig:conceptual_image}
  \end{minipage}
  \hfill
  \begin{minipage}{0.48\linewidth}
    \centering
    \includegraphics[width=1.\linewidth]{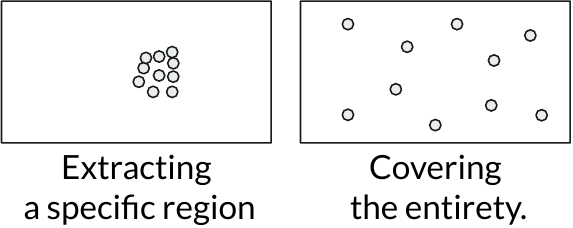}
    \vspace{-5mm}
    \subcaption{The core-set aims to cover an entire range, not a specific region.}
    \label{fig:diversity_image}
  \end{minipage}
  \vspace{-3mm}
  \caption{The purpose and characteristic of core-set selection.}
  \vspace{-5.5mm}
  \label{fig:key_visual}
\end{figure}

As a relevant machine learning technique, \textit{core-set selection} has been proposed~\cite{sener2017active}.
This method aims to extract a subset~(\textit{core-set}) that achieves an equivalent learning effect as the entire dataset, as shown in Fig.~\ref{fig:conceptual_image}.
Unlike \textit{point-wise data selection}, which considers each data point independently, this method is a kind of \textit{subset selection} and considers a subset as a whole when selecting data.
Furthermore, as shown in Fig.~\ref{fig:diversity_image}, it seeks to cover the entire range, rather than just extracting a specific region.

This paper proposes a core-set selection method for multi-speaker TTS. 
We define a \textit{diversity metric} based on language and speech features derived from self-supervised learning (SSL) models to assess the coverage area of a subset, formulating the core-set selection as a diversity maximization task under constraints on subset size.
Our proposed method is computationally lightweight. Specifically, it does not involve any model training with the entire dataset.
We conduct experiments on TTS corpora in Japanese, Chinese, and English to demonstrate that our core-set selection method mitigates the degradation in the naturalness and intelligibility of synthetic speech compared with phoneme balance based subset selection~\cite{kurematsu1990atr}.
The contributions of this work are as follows:
\begin{itemize} \vspace{-1.2mm} \itemsep -1mm \leftskip -5mm
    \item This paper is the first to introduce core-set selection in TTS tasks.
    \item Our proposed method is computationally efficient, mitigating degradation of naturalness and intelligibility in synthetic speech.
    \item We conduct experiments with multiple languages and dataset sizes to demonstrate the validity of our proposed method.
\end{itemize} \vspace{-1mm} 

\vspace{-2mm}
\section{Related work} 
\label{sec:related_work}
\vspace{-2mm}
\subsection{Designing phoneme-balanced speech corpora}
\label{sec:phoneme_balanced}
\vspace{-2mm}
A classical method for constructing a balanced speech dataset is to construct a phoneme-balanced sentence set~\cite{kurematsu1990atr} by maximizing the following function defined to evaluate the phoneme balance:
\begin{equation}
    H(\bm p) = -\sum_{i=1}^n p_i \log p_i
    \label{eq:entropy}
\end{equation}
where $p_i$ is the occurrence probabilities of the $i$-th phoneme and $n$ is the number of phonemes.
This method aims to enhance the effectiveness of model training by avoiding situations where a specific phoneme occurs extremely infrequently.
 
\vspace{-2mm}
\subsection{Data selection for multi-speaker TTS}
\vspace{-2mm}

Previous studies~\cite{lee2018comparison,seki2023} proposed data selection methods for multi-speaker TTS, but they were point-wise data selection methods whereas this study addresses a subset selection method.
A previous study~\cite{gallegos2020unsupervised} proposed a data subset selection method based on speaker selection, but that method extracts a specific region from the speaker distribution, whereas the subset selection method in this study aims to cover the entire range of the original dataset.
Previous studies~\cite{shamsi2019corpus, taubert2022comparison} proposed text set construction methods by clustering in the linguistic embedding space, and another~\cite{nose2017sentence} demonstrated a selection method based on phoneme and prosody entropy that is effective for statistical parametric speech synthesis.
However, they do not consider other aspects such as speaker identity, important factors for multi-speaker TTS.
In other words, core-set selection methods with the goal of efficiently training large-scale multi-speaker TTS models have not been explored.
    
\vspace{-3mm}
\subsection{Diversity-based subset selection algorithms}
\label{sec:relatet_work:diversity}
\vspace{-2mm}
A previous study~\cite{sener2017active} proposed a diversity-based core-set selection using the $k$-center method~\cite{kcenter}, and that study uses a similar algorithm.
Diversity-based subset selection is applied in various domains including recommendation systems~\cite{chen2018fast}, and several diversity evaluation metrics have been proposed.
One such metric~\cite{mmr} is used to drive the $k$-center algorithm.
We adopt a modified version of this metric~\cite{max_sum_diversity_origin}, where $\mathrm{max}$ is replaced with $\mathrm{sum}$.

\vspace{-3mm}
\section{Proposed method}
\vspace{-2mm}
We first extract utterance-level feature vectors that encompass linguistic, speaker, and acoustic features for each data point.
Using these feature vectors, we define a diversity metric and use the proposed core-set selection to maximize the diversity metric.

\vspace{-2mm}
\subsection{Feature extraction}
\label{sec:feature}
\vspace{-2mm}
For diversity-based core-set selection, we use feature vectors where similar data appear close together and dissimilar data appear far apart in feature space.
Each data point in multi-speaker TTS corpora consists of a pair of text, speaker identity, and speech, and we utilize the joint features of each aspect.

\textbf{Linguistic features $\bm{x}_{\text{linguistic}}$}:
We use sentence embeddings of texts, which are expected to separate data from different text domains.
Specifically, we average the output vector sequences from BERT~\cite{BERT} to obtain fixed-dimensional vectors, as detailed in~\cite{SentenceBERT}, and then normalize them to have a norm equal to $1$.
Although this method may lead to the loss of specific features of individual words, it is still effective to place similar data points in close proximity in the feature space.

\textbf{Speaker features $\bm{x}_{\text{speaker}}$}: 
We use continuous speaker representations like $x$-vectors~\cite{xvector} to incorporate speaker similarity into similarity calculations for the data.
We normalize $x$-vectors to have a norm equal to $1$. 
For multi-style TTS, our proposed method can be applied by using a pre-trained style encoder to extract style features.

\textbf{Acoustic features $\bm{x}_{\text{acoustic}}$}:
We average the output vector sequences of wav2vec~2.0~\cite{wav2vec2} and normalize their norms.
SSL features have demonstrated their effectiveness in speech recognition~\cite{wav2vec2}, and phonetic information is believed to be represented in their frame-level features.
Averaging these features is analogous to using dense vectors instead of one-hot vectors in time-frequency representations and is expected to contribute to expanding the phonetic coverage of the core-set.

Since linguistic and speaker features correspond to the input to the TTS model, we define input features $\bm{x}_{\text{ipnut}}$ by concatenating them, and define output features $\bm{x}_{\text{output}}$ as the acoustic feature.
Finally, we concatenate input and output features to obtain the joint features $\bm{x}_{\text{joint}}$ used for calculating diversity.
The relationship between the features is described in Eq.~\eqref{eq:feature}. 

\begin{equation}
    \bm{x}_{\text{input}} = \begin{bmatrix} \bm{x}_{\text{linguistic}} \\ \bm{x}_{\text{speaker}} \end{bmatrix},
    \;
    \bm{x}_{\text{output}} = \bm{x}_{\text{acoustic}},
    \;
    \bm{x}_{\text{joint}} = \begin{bmatrix} \bm{x}_{\text{input}} \\ \bm{x}_{\text{output}} \end{bmatrix}.
    \label{eq:feature}
\end{equation}

\begin{figure}[!t]
\vspace{-2mm}
\begin{algorithm}[H]
    \caption{Select a core-set $S$ from dataset $D$}
    \label{alg1}
    \begin{algorithmic}[1]
    \STATE $S \leftarrow \varnothing$
    \STATE $\bm x \sim U(D)$
    \WHILE{$T(S\cup \{\bm x\})\leq t_{\mathrm{max}}$}
    \STATE $S \leftarrow S\cup \{\bm x\}$
    \STATE $\bm x \leftarrow \textrm{argmax}_{\bm x\in D\setminus S} \sum_{\bm y\in S} \|\bm x - \bm y\|^2$
    \ENDWHILE
    \end{algorithmic}
\end{algorithm}
\vspace{-8mm}
\end{figure}

\vspace{-2mm}
\subsection{Diversity evaluation metric}
\vspace{-2mm}
Let $S$ and $D$ respectively denote a data subset and the entire dataset, $\bm x,\bm y\in \mathbb{R}^d$ denote feature vectors, and $\|\cdot\|, \mathrm{cossim}$ represent $L^2$ norm and cosine similarity.
To assess the diversity of $S$, previous studies~\cite{max_sum_diversity_origin} proposed calculating $\sum_{\bm x,\bm y\in S} d(\bm x,\bm y)$ where $d(\cdot,\cdot)$ is dissimilarity (e.g., $\|\cdot\|$ or $-\mathrm{cossim}$).
Since $V(S)$ takes a higher value when $S$ contains many dissimilar data pairs, $V(S)$ is anticipated to represent the spread of $S$ in the feature space.
When the norms of $\bm x$ and $\bm y$ are equal to $1$, the relationship $\|\bm x-\bm y\|^2 = 2 - 2\mathrm{cossim}(\bm x,\bm y)$ holds and squared distance has a close relationship with cosine similarity.
Therefore, we adopt squared Euclidean distance as a dissimilarity metric and evaluate diversity using the following function $V(S)$, the same as a previous study~\cite{meng2018scalable}
\begin{equation}
    V(S) := \sum_{x,y\in S}\|\bm x-\bm y\|^2.   \label{Eq:diversity}
\end{equation}

\vspace{-4mm}
\subsection{Core-set selection algorithm}
\vspace{-2mm}
We conduct core-set selection by solving the optimization problem that involves maximizing the diversity score $V(S)$ for subset $S$ subject to size constraints about $S$.
However, this optimization problem is a combinatorial optimization and can lead to explosive computational complexity.
Since this study focuses on scenarios with a large corpus and requires algorithms with low computational resources, we utilize a greedy algorithm.
Specifically, we execute the core-set selection procedure by sequentially adding the data that maximizes the diversity score until we reach the desired core-set size.

When adding each data, we select $\bm x$ to maximize $V(S\cup\{\bm x\})$, which can be expressed as the sum of $V(S)$ and $2 \times \sum_{\bm y\in S} \|\bm x-\bm y\|^2$.
Since $V(S)$ does not depend on $\bm x$, the core-set selection procedure follows the algorithm outlined in Algorithm~\ref{alg1}, where $U(D)$ represents a uniform distribution on $D$, $T(S)$ represents total speech duration included in $S$, and $t_{\mathrm{max}}$ represents a constraint on $T(S)$.
Notably, this algorithm can be executed with feature vectors in instead of the actual data, leading to reduced storage requirements.

\vspace{-2mm}
\section{Experimental evaluation} \vspace{-2.5mm}
\subsection{Experimental conditions} 
\vspace{-2mm}
\subsubsection{Dataset}
\label{sec:dataset}
\vspace{-2mm}
We trained a \textit{monolingual multi-speaker TTS model} using a multi-speaker TTS corpus for each of Japanese, Chinese, and English.
We used parallel~100 and nonparallel~30 subsets from the JVS~\cite{JVS} corpus for Japanese,
the AISHELL-3~\cite{aishell3} corpus for Chinese, 
and the training sets from the LibriTTS-R~\cite{LibriTTS-R} corpus (train\_clean\_100 and train\_clean\_360) for English.
The corpus sizes are $25$-hour, $63$-hour, and $243$-hour, respectively, from which the core-sets are selected.
Each corpus includes $100$, $174$, and $1151$ speakers. 

For text datasets for evaluating Japanese TTS models, we used $324$ sentences from the ITA corpus~\cite{ita_git}.
For the other languages, we randomly selected $100$ sentences from the test set of the corpora.

\begin{table}[t]\caption{Model-wise pseudo-MOS for TTS models trained with each dataset. The values in parentheses represent the core-set in hours. Bold indicates the highest value among the subsets.}
\label{table:mean_UTMOS}
\vspace{-2mm}
\centering \footnotesize
\begin{tabular}{c||c||c|c|c}
\multirow{2}{*}{Dataset} & All & Phoneme & Input & Our \\
 & Data & Balance & Balance & method \\ \midrule 
JVS~(3h) & $3.020$ & $2.971$ & $2.942$ & $\mathbf{2.996}$ \\
JVS~(6h) & $3.020$ & $2.966$ & $3.006$ & $\mathbf{3.080}$ \\
JVS~(12h) & $3.020$ & $3.022$ & $2.996$ & $\mathbf{3.080}$ \\
AISHELL-3~(6h) & $2.604$ & $2.717$ & $2.724$ & $\mathbf{2.796}$ \\
LibriTTS-R~(25h) & $3.943$ & $3.875$ & $3.891$ & $\mathbf{3.902}$ \\
\end{tabular}
\vspace{-1.5mm}
\end{table}

\begin{table}[t]\caption{ASR error rate~$(\%)$ in transcribing synthesized speech. A lower value implies higher intelligibility of synthesized speech. Bold indicates the lowest value among the subsets.}
\label{table:asr_result}
\vspace{-2mm}
\centering \footnotesize
\begin{tabular}{c||c||c|c|c}
\multirow{2}{*}{Dataset} & All & Phoneme & Input &  Our \\
 & Data & Balanced & Balanced & method \\ \midrule 
JVS~(3h) & $19.92$ & $24.53$ & $24.25$ & $\mathbf{21.26}$ \\
JVS~(6h) & $19.92$ & $23.37$ & $22.42$ & $\mathbf{20.25}$ \\
JVS~(12h) & $19.92$ & $21.19$ & $21.91$ & $\mathbf{20.37}$ \\
AISHELL-3~(6h) & $15.55$ & $18.09$ & $18.93$ & $\mathbf{17.97}$ \\
LibriTTS-R~(25h) & $16.61$ & $18.22$ & $\mathbf{17.54}$ & $17.84$ \\
\end{tabular}
\vspace{-1.5mm}
\end{table}

\vspace{-2mm}
\subsubsection{Model and training} 
\label{sec:ModelAndTraining}
\vspace{-2mm}
The multi-speaker TTS models included FastSpeech 2~\cite{FastSpeech2} and the pre-trained HiFi-GAN vocoder~\cite{Hifi-gan} UNIVERSAL\_V1~\cite{hifigan_git}.
We followed hyperparameters of the open-source implementations~\cite{FastSpeech2-JSUT,FastSpeech2_git}.
For speaker representation, we opted for $512$-dimensional $x$-vector, using a pre-trained model~\cite{jtube_xvector}. 
Each unique speaker corresponded to one $x$-vector, with all $x$-vectors having an L2 norm equal to $1$.
The $x$-vector was added to the output of the FastSpeech~2 encoder via a $512$-by-$256$ linear layer. 
The number of training steps was set in accordance with the size of each corpus: $50k$ steps for JVS and AISHELL-3, and $300k$ steps for LibriTTS-R.

\vspace{-2mm}
\subsubsection{Feature extractors for core-set selection} 
\vspace{-2mm}
We used language-specific BERT models~\cite{LINE_DistilBERT_Japanese,ChineseBERT,EnglishBERT} for linguistic features, 
$x$-vectors for speaker features,
and wav2vec 2.0~\cite{wav2vec2} base model~\cite{wav2vec_git} pre-trained on Librispeech~\cite{LibriSpeech} for acoustic features.

\vspace{-2mm}
\subsubsection{Compared methods} 
\vspace{-2mm}
\label{sec:比較手法}
We compared the following data selection methods.

\textbf{All Data}: 
To assess the quality degradation in training with subsets, we conducted training using the entire dataset.

\textbf{Phoneme Balance}: 
As a conventional method, a subset was selected to maximize phoneme entropy as described in Sec.~\ref{sec:phoneme_balanced}.

\textbf{Input Balance}: 
Expanding phoneme balance to multi-speaker scenarios, we used subset selection by maximizing the sum of phoneme entropy and speaker ID entropy, under the expectation of enhancing the learning effect by reducing speaker imbalance.

\textbf{Our method}:
Our diversity-based core-set selection.
We calculated the joint feature vector for each data point and then incrementally added data to the core-set, maximizing the diversity score.

\vspace{-2mm}
\subsubsection{Comparison conditions}
\vspace{-2mm}
We conducted experiments to answer the following questions:

\textbf{Does our method work?: validation with varying core-set size.}
To validate whether our method is more effective than traditional balance-based methods, experiments were conducted with multiple core-set size.
Core-sets of approximately $10, 20, 40\%$ of JVS were evaluated, corresponding to $3, 6, 12$ hours, respectively.

\textbf{Does our method work across language and corpus size?: validation on multiple corpora.}
We compared the balance-based methods and our method across varying languages and corpus sizes, specifically using AISHELL-3 and LibriTTS-R.
We selected $6$-hour and $25$-hour core sets, which correspond to about $10\%$ of the corpus.

\textbf{Are joint features effective?: ablation study about features.}
To assess the effectiveness of combination of input and output features, we conducted core-set selection with each feature. 
Core-set sizes were set to $3, 6, 12$ hours.
\vspace{-2mm}
\subsubsection{Evaluation criteria} \vspace{-2mm}
We synthesized speech for all speakers included in each corpus with test sentences prepared in Sec.~\ref{sec:dataset} and evaluate them using both automated and human subjective assessments.

For automatic evaluation, we used pseudo-MOS, an automatically predicted mean opinion score~(MOS) of synthetic speech.
Specifically, we used the UTMOS~\cite{UTMOS} strong learner model~\cite{UTMOS_git}, which has high accuracy in English, Chinese~\cite{huang22voicemos}, and Japanese~\cite{seki2023}.

\textbf{Model-wise pseudo-MOS evaluation:} To quantitatively compare the overall performance of multi-speaker TTS models, pseudo-MOSs were averaged over all speakers for each model.

\textbf{Speaker-wise pseudo-MOS evaluation:} To assess speaker-wise performance, the pseudo-MOSs were averaged per speaker.

We also calculated recognition error rates by using an automatic speech recognition~(ASR) model, Whisper~\cite{whisper} large model.
We evaluated character error rate for Japanese, phoneme error rate in Chinese, and word error rate in English, which are referred to as ASR error rate.

As a subjective evaluation experiment, we conducted MOS tests on speech naturalness and aggregated speaker-wise and model-wise MOS.
The evaluation for JVS encompassed all data, both balance methods for the $3$-hour core-set, and our method for the $3$, $6$, and $12$-hour core-set.
For the other corpora, all methods were included.
There were $800$ evaluators for JVS and $400$ for every other corpus.
Each listener assessed $24$ samples using a 5-point scale.
Since LibriTTS-R has a large number of speakers, we sampled speakers at intervals of $10$ in pseudo-MOS order and evaluated $116$ speakers.
We evaluated all speakers for the other corpora.

Furthermore, in an ablation study about features, we conducted subjective preference tests comparing each feature with joint features, using $3$-hour core-sets.
For each comparison, $100$ evaluators assessed the naturalness of synthesized speech for 10 randomly selected combinations of speakers and sentences. 

\begin{figure}[t]
\centering
\includegraphics[width=1.0\linewidth]{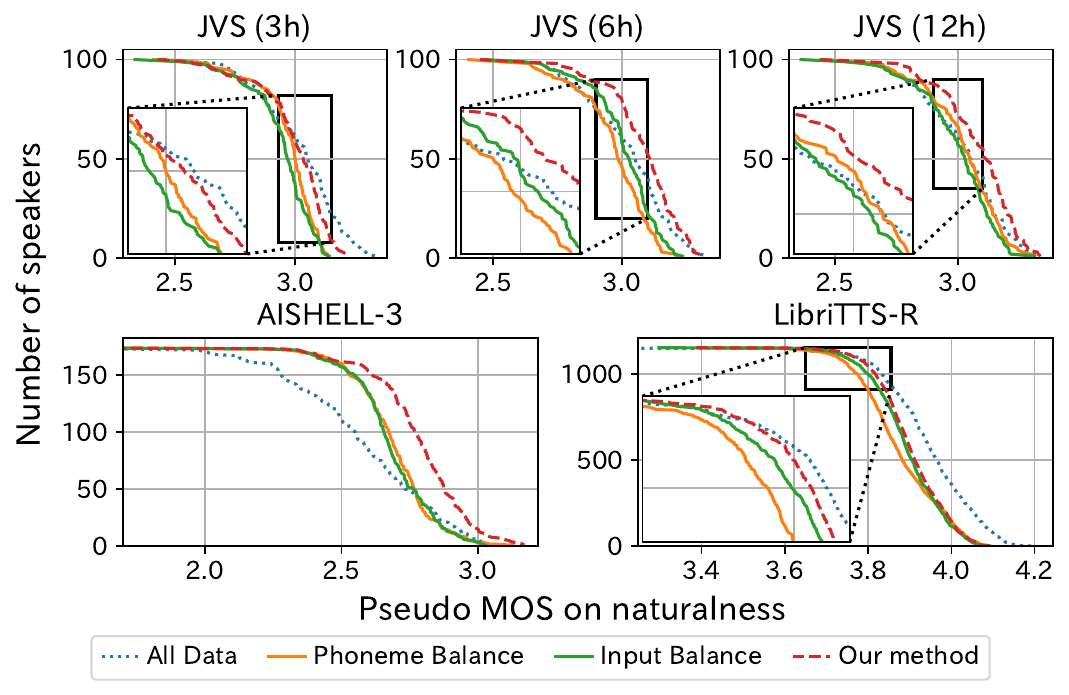}
\vspace{-7mm}
\caption{Cumulative histogram of speaker-wise pseudo-MOS for each model. The values on the $y$-axis represent the number of speakers with pseudo-MOS scores higher than the values on the $x$-axis.}
\vspace{-2.5mm}
\label{fig:all_columtive_histogram}
\end{figure}

\vspace{-2mm}
\subsection{Results}
\vspace{-2mm}
\subsubsection{Validation with varying core-set size}
\vspace{-2mm}
The upper half of Table~\ref{table:mean_UTMOS} presents the model-wise pseudo-MOSs for the models trained on core-sets from JVS.
Our method consistently outperforms the other balance methods across all core-set sizes, with an average improvement of $0.068$.
Considering a previous study~\cite{seki2023} that demonstrated that pseudo-MOS values exhibit approximately half the range of MOS variation in Japanese, we can expect a wider range of improvement in MOS. 
Additionally, in the speaker-wise pseudo-MOS evaluations shown in the upper part of Figure~\ref{fig:all_columtive_histogram}, our method consistently achieves higher scores than balance methods across most ranges.
Furthermore, our method has better ASR error rates than the other balance methods among all subset sizes, as shown in the upper half of Table~\ref{table:asr_result}.
These results demonstrate that our method works to mitigate the decrease in naturalness and intelligibility of synthetic speech regardless of core-set size.

\begin{figure}[t]
  \centering
  \begin{minipage}{1\linewidth}
    \centering
    \includegraphics[width=1.0\linewidth]{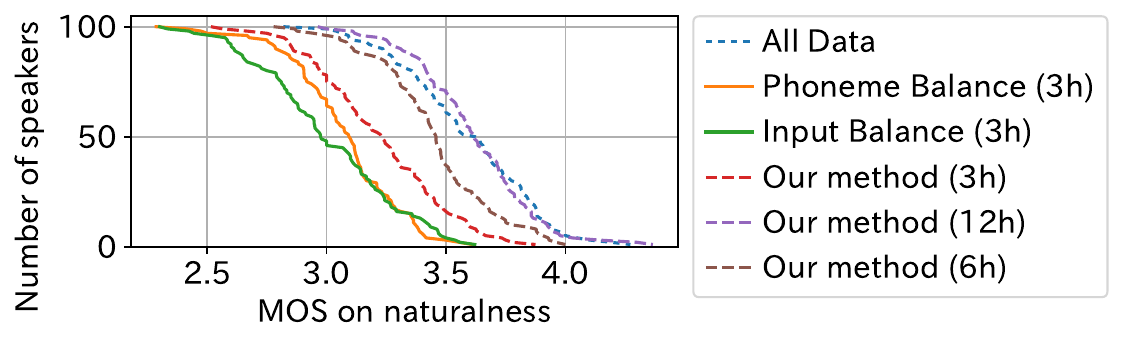}
    \vspace{-6mm}
    \subcaption{Comparison with the JVS corpus.}
    \label{fig:subjective_columtive_histogram}
  \end{minipage}
  \vfill
  \begin{minipage}{1\linewidth}
    \centering
    \includegraphics[width=1.0\linewidth]{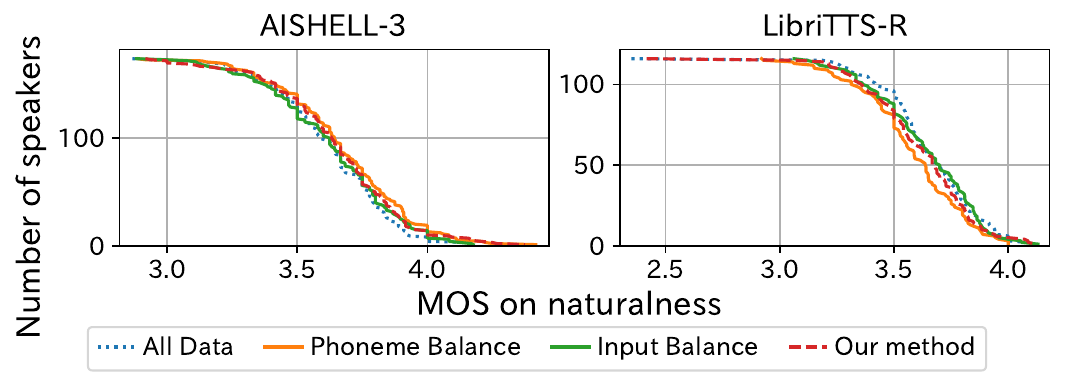}
    \vspace{-6mm}
    \subcaption{Comparison with the AISHELL-3 corpus and the LibriTTS-R corpus.}
    \label{fig:foreign_subjective_columtive_histogram}
  \end{minipage}
  \vspace{-2.5mm}
  \caption{Cumulative histogram of speaker-wise MOS for each model.}
  \vspace{-4.5mm}
\end{figure}

Figure~\ref{fig:subjective_columtive_histogram} illustrates the results of the subjective evaluation. 
Within the $3$-hour core-sets, our method consistently produces higher curves than the balance methods, indicating that our method improves naturalness of synthetic speech for all speakers.
The model-wise MOSs for phoneme balance, input balance, and our method were $3.069$, $2.999$, and $3.217$, respectively. 
Consequently, our method achieved an average improvement of $0.183$ in model-wise MOS, demonstrating its effectiveness in terms of naturalness in subjective evaluations.

Notably, the model-wise MOS with the $12$-hour core-set selected by our method is $3.614$, which closely matches the $3.587$ achieved by all data.
Also, this core-set achieved nearly equal values to all data in terms of pseudo-MOS and ASR error rates~(see Tables~\ref{table:mean_UTMOS},\ref{table:asr_result}).
These results indicate that the core-set attained an equivalent learning effect to that of the entire dataset in terms of naturalness and intelligibility.
In other words, it highlights the validity of using core-set selection as advantageous in terms of data volume.

\vspace{-2mm}
\subsubsection{Validation on multiple corpora}
\vspace{-2mm}
The lower half of Table~\ref{table:mean_UTMOS} displays model-wise pseudo-MOSs for Chinese and English.
In both AISHELL-3 and LibriTTS-R, our method outperforms the balance-based methods, demonstrating an average improvement of $0.101$ and $0.018$, respectively.
Furthermore, in terms of the increment of ASR error rates~(shown in the lower half of Table~\ref{table:asr_result}) compared with all data, the input balance exhibits $3.38\%$ in AISHELL-3 while the proposed method reduces to $2.42\%$.
In the case of LibriTTS-R, the phoneme balance method results in $1.61\%$, whereas the proposed method reduces to $1.23\%$.
From these results, we can say the TTS model trained with the core-set selected by the proposed method can mitigate degradation in naturalness and intelligibility better than the balance methods.

The lower half of Figure~\ref{fig:all_columtive_histogram} shows the results of speaker-wise pseudo-MOS in Chinese and English.
For AISHELL-3, our method's curve is shifted more to the right than the others, clearly indicating its superiority over the other balanced methods.
The lower performance of all data is attributed to its imbalance, e.g., phonemic imbalance; the phonemic entropy (Eq.~\eqref{eq:entropy}) dropped from $6.8$ in the phoneme balance method to $6.5$.
For LibriTTS-R, although the difference between our method and the other balance-based methods is marginal, the zoomed-in figure~(bottom left) reveals that our method has fewer speakers with low pseudo-MOSs.
This implies that, while other methods suffer a decrease in quality for speakers with less data, our method effectively addresses and corrects this issue.

The lower half of Figure~\ref{fig:foreign_subjective_columtive_histogram} shows the speaker-wise MOS in Chinese and English.
For AISHELL-3, the model-wise MOS for all data, phoneme balance, input balance, and our method were $3.627$, $3.679$, $3.634$, and $3.657$, respectively.
In the case of LibriTTS-R, the corresponding scores were $3.642$, $3.568$, $3.632$ and $3.606$.
MOSs for our method are nearly equivalent to those of all data for all speakers in both corpora, indicating that the proposed method performs at a level similar to that of the entire dataset in terms of MOS.

From these results, we conclude our method is applicable and effective irrespective of language or corpus size.

\begin{table}[t]\caption{Model-wise pseudo-MOS for each feature. Values in parentheses represent difference from all data.}
\label{table:feature_compare}
\vspace{-2mm}
\centering \footnotesize
\begin{tabular}{c||c||c|c|c}
Core-set & All & Input & Output & Joint \\
size & Data & features & features & features \\ \midrule 
$3$-hour & $3.020$ & $2.943 ({-0.077})$ & $2.931 ({-0.089})$ & $2.996 ({-0.024})$ \\
$6$-hour & $3.020$ & $3.032 ({+0.012})$ & $2.984 ({-0.036})$ & $3.080 ({+0.060})$ \\
$12$-hour & $3.020$ & $3.040 ({+0.020})$ & $3.047 ({+0.027})$ & $3.080 ({+0.060})$ \\
\end{tabular}
\vspace{-3mm}
\end{table}

\begin{table}[t]\caption{
    Subjective evaluation on the naturalness of synthesized speech. 
    Comparison between individual features and a joint feature.}
\label{table:ab_preference}
\vspace{-2.5mm}
\centering \footnotesize
\begin{tabular}{c|ccc|c}
\multirow{2}{*}{Compared} & \multicolumn{3}{c|}{Score} & \multirow{2}{*}{$p$~value} \\
 & Compared & vs. &  Joint features  &   \\ \midrule
Input features & $0.441$ & vs. & \;$\mathbf{0.559}$ & $9.82 × 10^{-5}$\\
Output features & $0.475$ & vs. & \;$0.525$ & $1.03 × 10^{-1}$\\
\end{tabular}
\vspace{-2mm}
\end{table}

\vspace{-2mm}
\subsubsection{Ablation study about features}
\vspace{-2mm}
Table~\ref{table:feature_compare} presents model-wise pseudo-MOSs.
Joint features exhibit higher values than the other features.
Particularly within the 3-hour core-sets, where the other features showed a decrease of $0.077$ and $0.089$, joint features reduced it to $0.024$.
Additionally, Table~\ref{table:ab_preference} presents the results of the subjective evaluation.
Although the $p$-value against output features is not very small, the results suggest that joint features work to reduce the degradation of synthetic speech naturalness.
From these results, we can say that combining input and output features is suitable for measuring similarity in the training data.

\vspace{-3mm}
\section{Conclusion}
\vspace{-2mm}
We proposed a core-set selection method for multi-speaker text-to-speech~(TTS), which extracts a diverse subset on the basis of language and acoustic features.
Experimental results demonstrated that our proposed method improves the learning effect compared with phoneme balance based subset selection across multiple languages, corpora, and core-set sizes.
We anticipate that our proposed method will remain applicable even for larger corpora, thanks to its low computational and storage requirements.
Our future work includes conducting empirical experiments with even larger corpora.

\newpage
\printbibliography

\end{document}